\documentclass[aps,showpacs,prb,floatfix,twocolumn,amsmath,citeautoscript,longbibliography]{revtex4-1}
\usepackage{graphicx}
\usepackage{amsmath}
\usepackage{amssymb}
\usepackage{bm}
\usepackage{dcolumn}
\usepackage{subfigure}
\usepackage{units}
\usepackage{hyperref}
\usepackage[usenames]{color}
\usepackage{booktabs}
\usepackage{multirow}
\usepackage{siunitx}
\usepackage{wrapfig}
\hypersetup{pdfborder=0 0 0,colorlinks=true,citecolor=blue,linkcolor=blue}
\usepackage{natbib}

\newcommand{\MX}{MX$_2$}
\newcommand{\MS}{MoS$_2$}

\begin{document}

\title{One-dimensional electronic instabilities at the edges of \MS}%

\author{Sridevi Krishnamurthi, Mojtaba Farmanbar, and Geert Brocks}%
\email[]{g.h.l.a.brocks@utwente.nl}
\affiliation{Computational Materials Science, Faculty of Science and Technology and MESA+ Institute for Nanotechnology, University of Twente, the Netherlands. }
\date{\today}

\begin{abstract}
The one-dimensional metallic states that appear at the zigzag edges of semiconducting two-dimensional transition metal di-chalcogenides (TMDCs) result from the intrinsic electric polarization in these materials, which for D$_{3h}$ symmetry is a topological invariant. These 1D states are susceptible to electronic and structural perturbations that triple the period along the edge. In this paper we study possible spin density waves (SDWs) and charge-density waves (CDWs) at the zigzag edges of {\MS}, using first-principles density functional theory calculations. Depending on the detailed structures and termination of the edges, we observe either combined SDW/CDWs or pure CDWs, along with structural distortions. In all cases the driving force is the opening of a band gap at the edge. The analysis should hold for all group VI TMDCs with the same basic structure as {\MS}.
\end{abstract}

\maketitle

\section{Introduction}

Transition metal di-chalcogenides (TMDCs) have emerged in the last decade as a new class of two-dimensional (2D) materials with attractive electronic and optical properties. In particular, the group of compounds {\MX}, M = Mo, W; X = S, Se, Te, has been at the center of attention. In the bulk $2H$-phase, these materials are indirect semiconductors. Because they have a layered structure, where the interlayer bonding is Vanderwaals, two-dimensional (2D) layers can be isolated by micromechanical cleaving, or they can be grown by chemical vapor deposition (CVD) or atomic layer deposition (ALD). These 2D layers are direct semiconductors with band gaps in the range 1-2 eV\cite{PhysRevLett.105.136805,Splendiani}, which makes them interesting for optoelectronic applications, and, as they also exhibit some catalytic activity, for photo-catalytic applications.

Growing 2D {\MX} layers, one naturally produces finite-sized structures with edges and grain boundaries\cite{PhysRevLett.87.196803,Ma}. Somewhat surprisingly, such one-dimensional (1D) structures are typically not semiconducting, but metallic at room temperature. To find edges to be metallic is not so extraordinary as such, as the atoms at the edge lack the full coordination of atoms in the bulk. Hence they commonly have dangling bonds that are partially occupied, leading to metallicity. This type of metallicity is however fragile, as it vanishes when the dangling bonds are saturated by adsorbants, which readily happens under ambient conditions, for instance. In contrast, the metallicity of {\MX} edges and grain boundaries seems to be robust. Not only does it occur in ultra-high vacuum (UHV), but also under ambient conditions, and it does not seem to depend critically on the details of the edge structure\cite{acs.nanolett.6b04715,acs.nanolett.7b02192,Barja:2016aa}.

The 1D edges and grain boundaries of 2D {\MX} layers seem therefore well suited to study the physics of 1D metallic systems. As both electron-electron and electron-lattice interactions are particularly effective in one dimension, one expects 1D metals to be very responsive to such interactions. From experiment, different scenarios for such responses have been proposed, ranging from Peierls structural reconstructions or charge density waves (CDWs) driving a metal-insulator transition, to a Tomonaga-L\"{u}ttinger liquid with spin-charge separation\cite{Barja:2016aa,PhysRevX.9.011055,Ma:2017aa}. 

Several computational studies have addressed the electronic structure of {\MX} edge states, of zigzag edges in particular. The emergence of magnetic moments on the metal atoms at the edges has been proposed \cite{PhysRevB.67.085410,Giulia,Botello_M_ndez_2009,Li,PhysRevB.80.125416}, and reconstructions of the {\MX} edges have been explored, where, depending on the exact edge termination, structures have been suggested that triple the period along the edge\cite{doi:10.1021/acs.chemmater.5b00398}, or double the period \cite{cui}. Some terminations are found to give rise to multiple (meta)stable structures\cite{doi:10.1002/chem.202000399}. In previous work, we have argued by means of first-principles calculations, that in mirror twin boundaries (MTBs), spin density waves (SDWs) coupled with charge density waves (CDWs), localized at the MTBs, drive a metal-insulator transition, with band gaps $\sim 0.1$ eV as a result\cite{krishnamurthi2020}, where the symmetry of the {\MX} lattice dictates the periodicity of these SDW/CDWs.   

In this paper we generalize these results by studying possible SDW/CDWs at different edge structures, using first-principles calculations at the level of density functional theory (DFT), including DFT+U. We use monolayer {\MS} as an example of the class of {\MX} materials. The edges with zigzag orientation are the most interesting, because they are the ones that emerge naturally under growth conditions, so we focus on these. We argue that at these edges, possible SDW/CDWs should have a periodicity of (a multiple of) $3a$, with $a$ the primitive lattice constant along the edge, which is dictated by topology and the symmetry of the lattice.

This paper is organised as follows. In Sec. \ref{sec:methods}, we present our structural nanoribbon model for the edges, and the technical details of the DFT calculations. We also summarize the topological and symmetry arguments that predict metallic edge states, which are partially occupied states localized at the edges, with energies within the band gap of the {\MX} monolayer, as well as the exact occupancy dictated by these arguments. In Sec. \ref{sec:results}, we present the results on the different zigzag edges, the Mo edge and the S edge, and of different structural and chemical modifications of these edges. Finally, the results and conclusions are summarized in Sec. \ref{sec:summary}.

\section{Methods and model}\label{sec:methods}
We use first-principles electronic structure calculations at the level of density functional theory (DFT). All DFT calculations are done with the VASP package\cite{PhysRevB.54.11169,PhysRevB.59.1758,PhysRevB.50.17953,PhysRev.140.A1133,PhysRevB.23.5048}, using the GGA/PBE and PBE+U functionals and the projected augmented wave (PAW) method, treating the Mo $4d$, $5s$, $4p$, and the S $3p$ and $3s$ shells as valence electrons. We use a plane wave kinetic energy cut-off of 400 eV, and a $k$-point sampling along the edge of 24 and 8 points for cells with $1\times$, and $3\times$ periodicity, respectively. Upon relaxing the structures, the ultimate forces acting on the atoms are below 0.01 eV/\AA, and the electronic convergence criterion is set to $10^{-5}$ eV/cell.

Although on-site Coulomb and exchange interactions in 4d transition metals, such as Mo, are typically weaker than in 3d transition metals, first-principles calculations on transition metals and transition metal oxides have shown that Hubbard $U$ values of order 3 eV are quite reasonable for 4d transition metal atoms \cite{PhysRevB.86.165105,PhysRevB.83.121101}. In 2D materials, where electronic screening is weaker than in 3D, one may expect similar values at least. We include such on-site Coulomb and exchange interactions within the GGA+U formalism, applying the rotational average approach \cite{PhysRevB.57.1505}, which uses a single parameter $U-J$, where we have checked $U-J$ values over a range from 2 eV to 4 eV. Including the on-site interaction with this parameter setting, does not change the electronic structure of the semiconducting monolayer {\MS}, but it can modify the (electronic) structure of (near-)metallic edges. 

We model the {\MS} edges by a nanoribbon structure, which is periodic in one dimension (the $x$-direction in the following). The width of the ribbon is 11 {\MS} formula units (the $y$-direction), and its height (the $z$-direction) is one unit. With this width the two edges of the ribbon do not interact directly. For too small a width ($\lesssim 3$ formula units) a bonding/anti-bonding interaction between the two edges occurs. The periodic images of the nanoribbon are separated from one another by a vacuum space of 15 \AA\  along both $y$- and $z$-directions.

It is important to note that in DFT calculations, if one uses a nanoribbon geometry with two different edges, and both the edges are metallic, there can be an electron transfer between the edges in order to equilibrate the Fermi level. This electron transfer is spurious if the nanoribbon is supposed to model  single edges as they occur under experimental conditions. It leads to a different electron occupancy at the two edges, and, in some cases, this may cause a reconstruction or CDW pattern that is restricted to a specific nanoribbon geometry \cite{cui}. To truly observe the intrinsic electronic properties of a single edge, the opposing edge of the nanoribbon must be made insulating to prevent this self-doping.

\subsection{Edge structures and metallicity}

\begin {figure}[htbp]
\includegraphics[width=8.5cm]{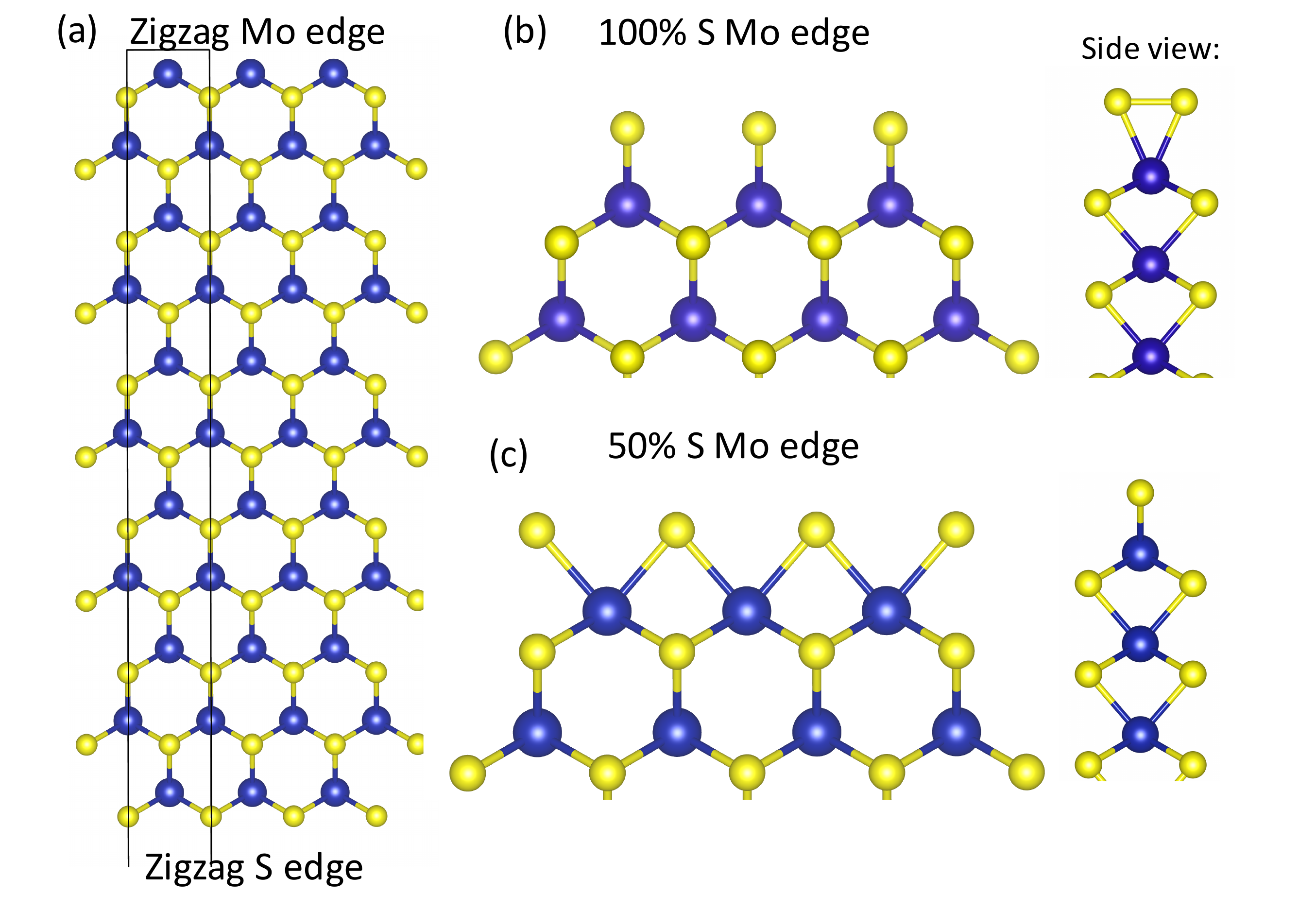}
\caption{ Structures of nanoribbons, with Mo (10) and S ($\bar{1}$0) zigzag edges along the $x$ direction. The dashed rectangle indicates $1\times$ unit cell and the blue and yellow spheres represent Mo and S atoms respectively. (a) Pristine Mo edge and S edge; (b) 100\% S-dressed Mo edge; (c) 50\% S-dressed Mo edge. }
\label{fig:structures}
\end{figure}

The elementary edges with (10) orientation of a 2D hexagonal lattice such as monolayer {\MS},  are called zigzag edges, see Fig. \ref{fig:structures}.  As the {\MS} lattice lacks inversion symmetry, or an in-plane twofold rotational or mirror symmetry, it means that for a nanoribbon with edges in zigzag orientation, the two opposite, (10) and ($\bar{1}$0), edges are structurally different. The (10) edge is called the Mo edge, as in its pristine form it is terminated by metal atoms, see Fig. \ref{fig:structures}(a). Similarly, the ($\bar{1}$0) edge is called the S edge; it is terminated by chalcogen atoms.\footnote{Edges with (12) orientation, which are perpendicular to the zigzag edges are called armchair edges. A nanoribbon with edges in armchair orientation has two identical edges. Armchair edges are semiconducting with a sizeable gap \cite{Li,doi:10.1021/acs.nanolett.5b02834}, and not of interest for the present paper.}

In a nanoribbon geometry, the $D_{3h}$ symmetry of monolayer {\MS} is obviously broken at the edges. Upon optimization, all structures we are considering here, however, turn out to keep the mirror symmetry $\sigma_{h}$ with respect to a plane through the Mo atoms, and consequently all electronic states can be classified as even or odd with respect to $\sigma_{h}$. 

Because of the lack of inversion or similar twofold symmetry, the polarization of the {\MS} lattice need not be zero. According to the modern theory of polarization, the latter can be calculated as\cite{PhysRevB.47.1651}
\begin{equation}
\mathbf{P}= \frac{e^2}{4\pi^2} \int_{BZ} \mathbf{A(k)}\ d^2k
\end{equation}
where $\mathbf{A(k)} = i\sum_{n,occ} \left\langle u_n(\mathbf{k})\left\vert\nabla_{k}\right\vert u_n(\textbf{k})\right\rangle$ is the Berry connection, with $u_n(\textbf{k})$ the periodic part of the Bloch wave function, and $n$ the band index. In 2D, the polarization vectors form a lattice $\mathbf{P}=p_1\mathbf{a}_1+p_2\mathbf{a}_2$, with $\mathbf{a}_1$ and $\mathbf{a}_2$ the 2D lattice vectors, and $(p_1,p_2)=(\alpha,\beta)+(n_1,n_2); n_{1,2}=0, \pm 1,\pm 2 ...$. For structures with $D_{3h}$ symmetry, the allowed values of $(\alpha,\beta)$ are restricted to $(0,0)$, $(1/3,2/3)$, or $(2/3,1/3)$, and the polarization is a topological $\mathbb{Z}_3$ invariant\cite{PhysRevB.88.085110,PhysRevB.86.115112}. Monolayer TMDCs, {\MX} (M = W, Mo, X = S, Se, Te) with $2H$ structure, have $(\alpha,\beta) = (2/3,1/3)$, as is confirmed directly by DFT calculations.

The nonzero polarization of a 2D {\MS} layer has direct consequences for the metallicity of its zigzag edges\cite{PhysRevB.87.205423,doi:10.1021/acs.nanolett.5b02834}. Formally it leads to a polarization charge $\lambda = \mathbf{P}\cdot\hat{\mathbf{n}} = \pm 2e/3a$ at a zigzag edge, where the positive and negative signs refer to zigzag edges of ($\bar{1}$0) and (10) orientations, respectively. In a macroscopic sample, the edges must be charge neutral, such as to avoid a polarization catastrophy. This means that the polarization charge has to be compensated by an electronic charge $-\lambda$. In some studies it is argued that the electronic charge resides in bands that are pulled up from the monolayer bulk valence band, or pushed down from the conduction band, due to a locally changed potential\cite{doi:10.1021/acs.nanolett.5b02834}. In the cases we have studied, we observe that the electronic charge fills additional states created at the edges, with energies inside the monolayer band gap. Such edge states are likely to have a topological origin; they are, for instance, also observed in tight-binding models\cite{PhysRevB.93.205444,PhysRevB.99.155109}, where the local potential does not change, but kept fixed at its bulk value.   

As the electrons residing in these edge states must contribute a charge $-\lambda$, it means that at the Mo edge (10), the corresponding 1D edge bands must have a total occupancy of $2/3$. Similarly, the corresponding 1D edge bands  at the S edge ($\bar{1}$0) must be $2/3$ occupied by holes. Clearly,  in absence of reconstructions or other 1D instabilities, both these edges have to be metallic. Moreover, from the occupancy of their edge states, one may suspect that the edges are susceptible to electronic and/or structural instabilities typical for 1D systems that triple the period along the edges.

\section{Results and discussion}\label{sec:results}

We first examine the bare Mo edge with unperturbed $1\times$ periodicity, Fig.  \ref{fig:structures}(a), and then look at different terminations with S atoms at the Mo edge, Figs. \ref{fig:structures}(b) and (c) \cite{PhysRevB.96.165436}. Bare Mo edges can only exist under extreme growth conditions, or are made artificially under electron bombardment, for instance \cite{acs.nanolett.7b02192}, as any appreciable presence of sulfur dresses the Mo edge with S atoms\cite{PhysRevLett.84.951}. Of those, dressing with S dimers, Fig. \ref{fig:structures}(b), or with individual S atoms, Fig. \ref{fig:structures}(c) are the most common, each of the two structures thermodynamically being stable in a different range of S chemical potential. They are called the 50\% S-dressed Mo edge, and the 100\% S-dressed Mo edge in the following, respectively.

We study the electronic structures of the bare, the 50\% S-dressed, and the 100\% S-dressed Mo edges, initially in their unperturbed $1\times$ periodicity, and then look at possible instabilities that triple the period. Of these, reconstructions or Peierls distortions, which perturb the edge structure, are the most basic, but we will also investigate spin density and charge density waves that are mainly electronic. 

Under typical conditions {\MS} grows in the form of triangular islands, where the Mo edge, either in its 50\% or in its 100\% S-dressed form\cite{PhysRevLett.87.196803, PhysRevB.96.165436}, is the most prevalent edge termination. For completeness, we also discuss the possible structures and electronic structures of S edges and their susceptibility to (electronic) perturbations that triple the period. 

As stated in the previous section, in a calculation that uses a nanoribbon geometry with two different edges, there can be an electron transfer between the edges in order to equilibrate the Fermi level, which can lead to an electron occupancy different from $\pm 2/3$ at the two edges. Therefore, in order to model the intrinsic properties of a single edge, the opposing edge of the nanoribbon is then made insulating.

\subsection{Pristine edges}

\begin {figure}[htbp]
\includegraphics[width=8.5cm]{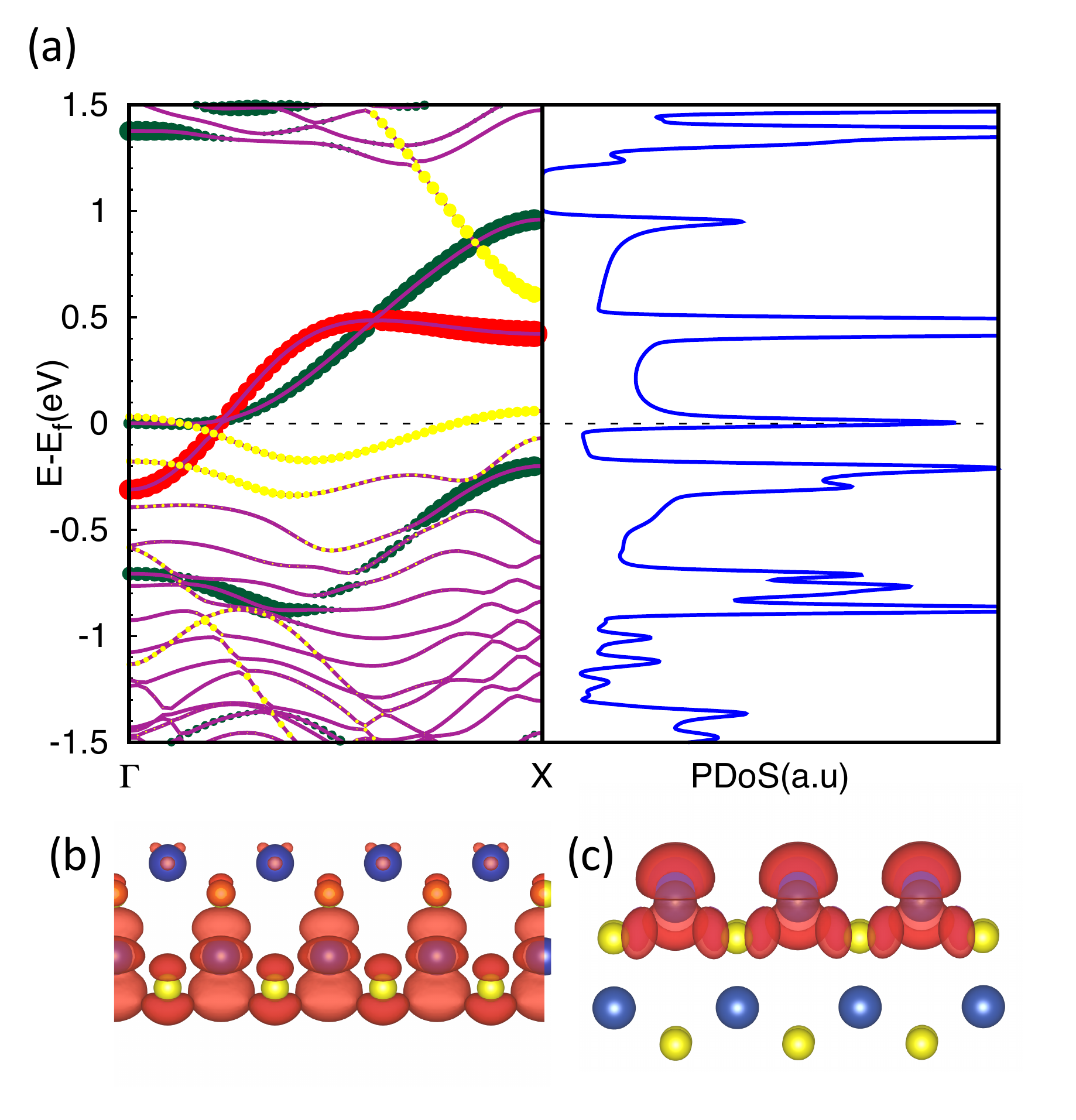}
\caption{ (a) Band structure of the pristine nanoribbon, calculated with the PBE functional; the green and red colors measure a projection of the wave function on the Mo orbitals of even ($d_{xy},d_{x^2-y^2},d_{3z^2-r^2}$), respectively odd ($d_{xz},d_{yz}$) $\sigma_h$ symmetry of the atoms at the Mo edge; the yellow color marks a similar projection on the $p$ orbitals of the S atoms at the S edge. The figure on the right side shows the projected density of states (PDoS), projected on the Mo atoms at the Mo edge. (b) Wave function density at the S edge at $E-E_F=-0.1$ eV. (c) Wave function density at the Mo edge at $E-E_F=0.5$ eV.}
\label{fig:pristineMoedge}
\end{figure}

A pristine Mo edge is terminated by Mo atoms, Fig. \ref{fig:structures}(a). The Mo atoms at the edge are undercoordinated and can be expected to participate in edge states. Indeed, the electronic structure of the nanoribbon shows two such dominantly Mo edge bands crossing the Fermi level. The first band highlighted in green in Fig. \ref{fig:pristineMoedge}(a), has Mo $d_{xy},d_{x^2-y^2},d_{3z^2-r^2}$ character and even symmetry with respect to $\sigma_h$, whereas the second band, highlighted in red, has Mo $d_{xz},d_{yz}$ character and odd symmetry. The latter band is shown below to have dangling bond character, and is sensitive to (chemical) changes, but the first band is more robust. The projected density of states (PDoS) at the Mo edge highlights the typical Vanhove singularities of 1D states, see Fig. \ref{fig:pristineMoedge}(a). 

A pristine S edge, terminated by S atoms, shows two edge bands. The first one, highlighted in yellow in Fig. \ref{fig:pristineMoedge}(a), crosses the Fermi level with very little dispersion. This band has S $p_{y},p_{z}$ character and odd symmetry, and can be classified as a typical dangling bond state. The second band is fully unoccupied, and disperses into the conduction band of \MS. It has  Mo $d$ character and even symmetry. Typical wave function densities for the Mo dominated states at the S edge and the Mo edge are shown in Fig. \ref{fig:pristineMoedge}(b) and (c).

 \subsection{100\% S-dressed Mo Edge}
 
At a Mo edge that is maximally dressed by S atoms, each of the Mo atoms at the edge is fully coordinated by six S atoms, see Fig. \ref{fig:structures}(b). This structure emerges under sulfur-rich growth conditions\cite{Flemming,Lauritsen,PhysRevB.67.085410,doi:10.1021/acs.chemmater.5b00398,PhysRevB.96.165436}. The S atoms at the edge have an incomplete coordination, which is partially offset by dimerization. In fact, the undimerized form is not stable, and S dimers form spontaneously at the edge during geometry optimisation.

\begin {figure}[htbp]
\includegraphics[width=8.5cm]{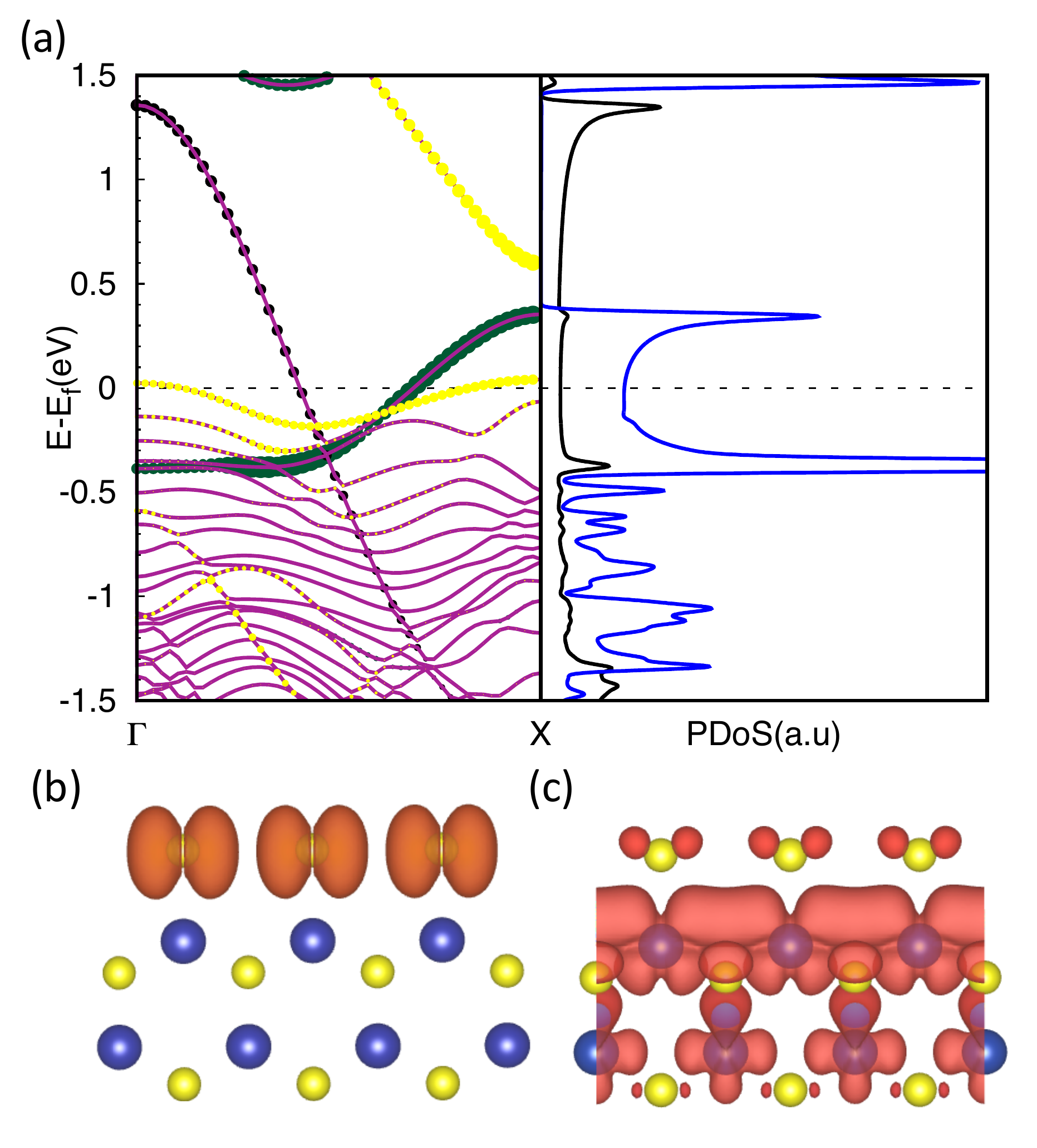}
\caption{(a) Band structure of nanoribbon with 100\% S-dressed Mo edge and corresponding PDoS in the same representation as in Fig.~\ref{fig:pristineMoedge}, calculated with the PBE functional. (b,c) Wave function densities at the Mo edge at $E-E_F=1.0$ eV and 0.0 eV, respectively.}
\label{fig:100dressededge}
\end{figure}

The band structure of the 100\% S-dressed edge is shown in Fig. \ref{fig:100dressededge}(a). The Mo edge remains metallic, with two edge bands crossing the Fermi level. A band with a large dispersion, downward from $\Gamma$ to X, toward the bulk states, can be attributed to the S dimers at the edge. It has S $p_{x}$ character and is completely localized on S atoms, see Fig.\ref{fig:100dressededge}(b). 

The band highlighted in green has Mo $d$ character and even $\sigma_h$ symmetry. It is in fact similar to the band with even symmetry identified in the pristine case, Fig. \ref{fig:pristineMoedge}(a), but lowered in energy by $\sim 0.5$ eV. Plotting the wave function density confirms its character, see Fig.\ref{fig:100dressededge}(c). The state is characterized by a bonding interaction between $d_{xy}$ orbitals of neighbouring Mo atoms, and an anti-bonding interaction with $p_{y}$ orbitals on the second row of S atoms and the $d_{x^2-y^2}$ orbitals of the second row of Mo atoms. The qualitative similarity to the state identified for the pristine edge, Fig. \ref{fig:pristineMoedge}(c), demonstrates the robustness of this state.

In contrast, the Mo-derived edge band with odd $\sigma_h$ symmetry identified for the pristine edge, (red color in Fig. \ref{fig:pristineMoedge}(a)) has disappeared for the S-dressed Mo edge, confirming our previous notion that it is a dangling bond state, which is sensitive to chemical perturbations.

Focusing on the occupation of the edge bands, we notice that the Mo $d$ band has an occupancy of approximately $2/3$, which one can also infer from the PDoS shown in Fig.\ref{fig:100dressededge}(a). This $2/3$ occupancy stems from the need for electronic compensation of the polarization charge, as discussed in the previous section.  

 \subsection{CDW/SDWs at the 100\% S-dressed Mo Edge}
 
Given the metallicity of the Mo edge, and the known susceptibility of 1D structures to electronic and structural perturbations, it is worth while to study whether such perturbations can break the translational symmetry along the edge. In particular, given the $2/3$ occupancy of the edge states, a Peierls type structural distortion may induce a metal-insulator transition at the edge in a $3\times$ unit cell. A similar transition may be induced by a charge density wave (CDW) or a spin density wave (SDW) with $3\times$ periodicity.

\begin {figure}[htpb]
\includegraphics[width=8.5cm]{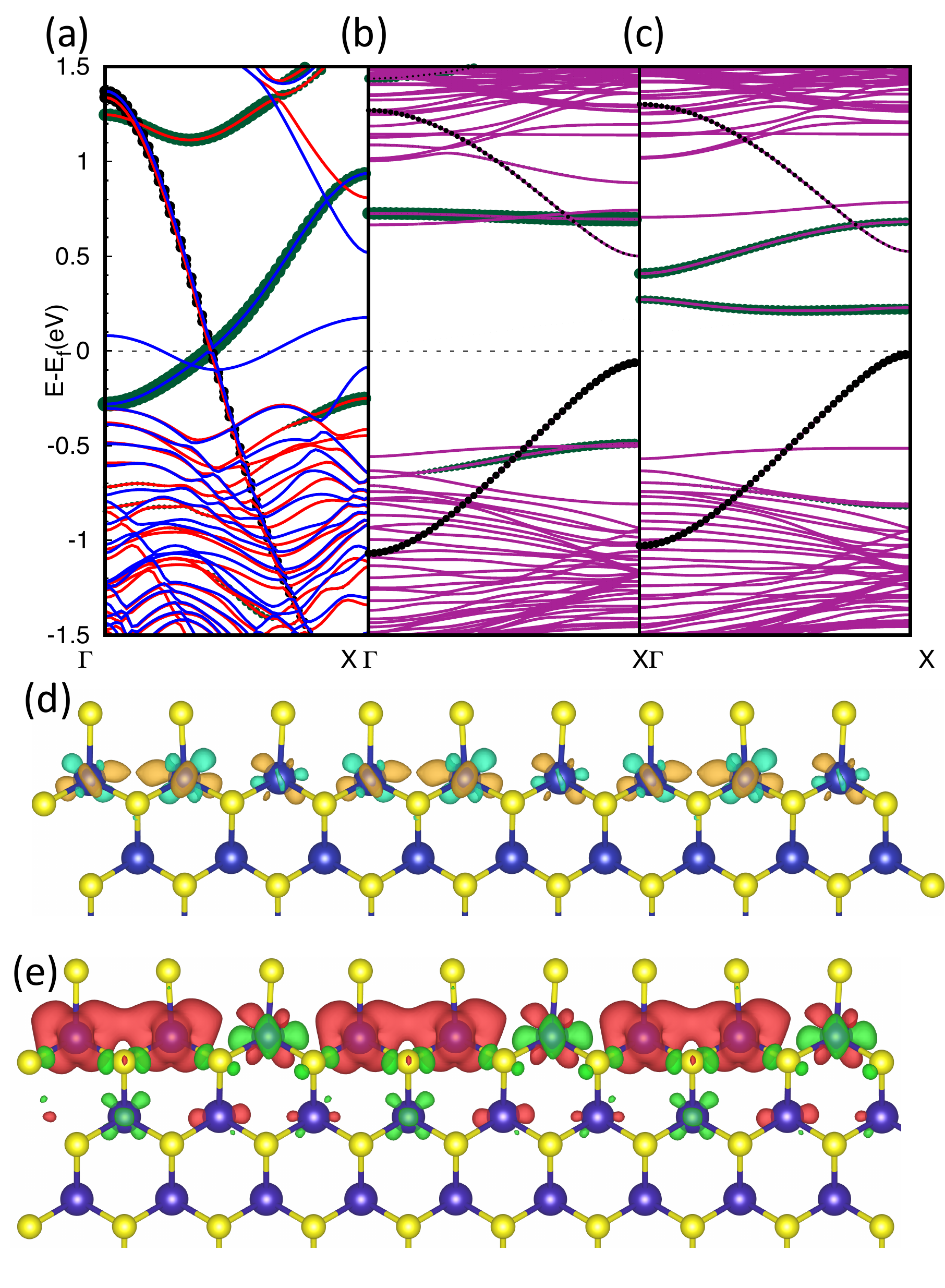}
\caption{ (a) Band structure of $1\times$ nanoribbon with 100\% S-dressed Mo edge, calculated with the PBE+U functional, $U-J=3$ eV, spin-up and down bands shown in the same figure in colors red and blue respectively. (b,c) Spin-up, spin-down bands in the $3\times$ periodic cell, after electronic and structural relaxation. (d) CDW at the Mo edge in this cell; the brown/green colors indicate the change in charge density $\Delta \rho = \rho_{3\times}-\rho_{1\times}$ with respect to that of the unperturbed $1\times$ structure. (e) Corresponding SDW at the Mo edge; the red/green colors indicate the density of spin up/down wave functions\cite{vesta}. }
\label{fig:100dressededgeSDW}
\end{figure}

Using the GGA/PBE functional, we do not find a Peierls distortion for the 100\% S-dressed Mo Edge, which is in agreement with the results obtained in previous work\cite{doi:10.1021/acs.chemmater.5b00398}. Including spin polarization at the GGA/PBE level does not change this result; it does not lead to any magnetic moments on the edge Mo atoms, for instance \cite{doi:10.1021/acs.chemmater.5b00398}. To study the possibility of SDWs at the edge, we therefore introduce a moderate $U-J = 3$ eV for the Mo 4d electrons\cite{PhysRevB.86.165105,PhysRevB.83.121101,PhysRevB.82.195128}. This addition does not change the electronic structure of a {\MS} monolayer, but it modifies that of the Mo edge. Starting with the structure with $1\times$ periodicity, the edge Mo atom develops a sizeable magnetic moment of 0.7$ \mu_{B}$. The Mo d edge band becomes spin split, with an energy splitting $\sim 1.0$ eV. The spin-up band becomes completely occupied and is pushed down into the valence band, whereas the spin-down band is pushed up, and becomes $1/3$ occupied, compare Figs. \ref{fig:100dressededgeSDW}(a) and \ref{fig:100dressededge}(a). In contrast, the band derived from the S $p_{x}$ states shows very little spin splitting, consistent with the absence of magnetic moments on the S atoms. 


The $1/3$ occupancy of the Mo edge spin-down band suggest to study possible perturbations of this 1D metallicity in a cell with $3\times$ periodicity  Indeed, upon reoptimization in this cell, a SDW emerges that breaks the $1\times$ periodicity, with magnetic moments on the three Mo atoms at the edge of $1.0$ $\mu_{B}$, $1.1$ $\mu_{B}$, and $-0.14$ $\mu_{B}$, respectively, see Fig. \ref{fig:100dressededge}(e). The SDW also gives rise to a modulation of the structure, accompanied by a CDW, Fig. \ref{fig:100dressededge}(d). Focusing on the structural details, the S atoms remain dimerized, where the distance between two adjacent S dimers displays the $3\times$ periodicity with values of 3.47 \AA\ and 2.97 \AA.  Likewise, the distances between adjacent edge Mo atoms also show this $3\times$ pattern with values  of 3.22 \AA\ and 3.09 \AA\ . The SDW/CDW causes a lowering of the total energy by 125 meV/$3\times$ unit cell. 

The corresponding band structures of the Mo edge clearly demonstrate that the emergence of the SDW/CDW leads to a metal-insulator transition. Folding the $1/3$ occupied Mo band of the $1\times$ cell in the $3\times$ cell gives three bands, with the lowest occupied and the upper two unoccupied, see \ref{fig:100dressededgeSDW}(b)(c). The SDW/CDW creates a sizeable band gap of $\sim 0.7$ eV between the occupied and unoccupied Mo edge bands (highlighted in green). The CDW also splits the S edge band, creating a gap of $\sim 0.5$ eV between occupied and unoccupied states (highlighted in black). The overall result is a band structure with an indirect gap of $\sim 0.2$ eV between a Mo edge state and a S edge state, \ref{fig:100dressededgeSDW}(c).

\begin {figure}[htpb]
\includegraphics[width=7.5cm]{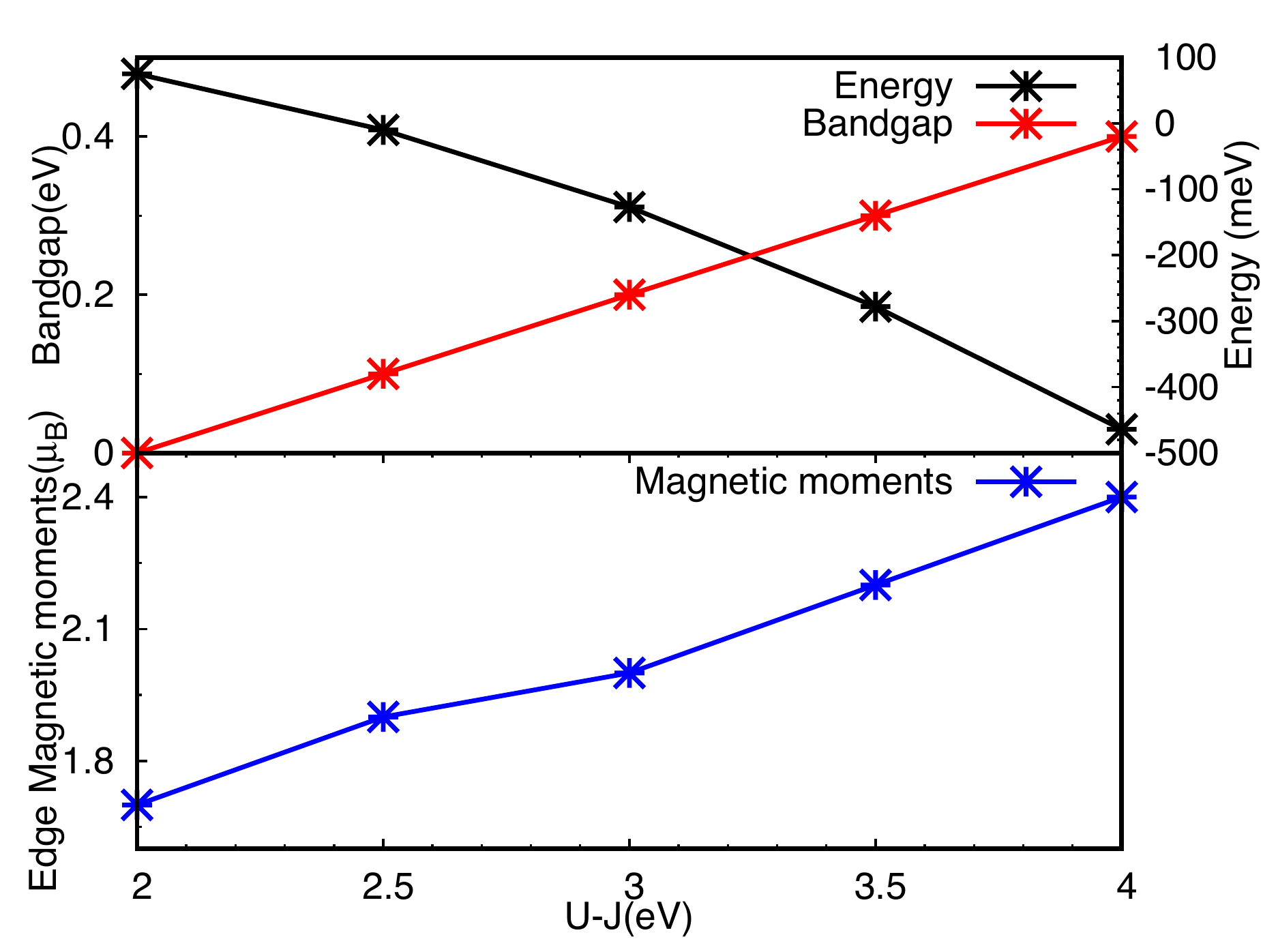}    
\caption{ Top: (black) total energy per $3\times$ cell of the SDW/CDW with respect to the non-polarized $3\times $ structure, as function of $U-J$; (red) corresponding band gap at the edge. Bottom: corresponding magnetic moments on the edge Mo atoms.}
\label{fig:U}
\end{figure}

The emergence of a SDW/CDW does not depend critically on the particular value of $U-J$ used in the calculation. It persists for $U-J > 2$ eV, although the band gap and the energy gain per unit cell increase monotonically with increasing $U-J$, see Fig.~\ref{fig:U}. Varying $U-J$ between 2 and 4 eV changes the magnetic moments on the Mo edge atoms by $\sim 50$\%. However, increasing $U-J$ to beyond 4 eV starts to spin-polarize all the ribbon states, not only the edge states, which seems unphysical. Through Bader analysis we do not find any evidence for multi-valency of the Mo edge atoms \cite{doi:10.1021/acs.chemmater.5b00398}.

\subsection{50\% S-dressed Mo Edge}
 
\begin {figure}[htpb]
\includegraphics[width=8.5cm]{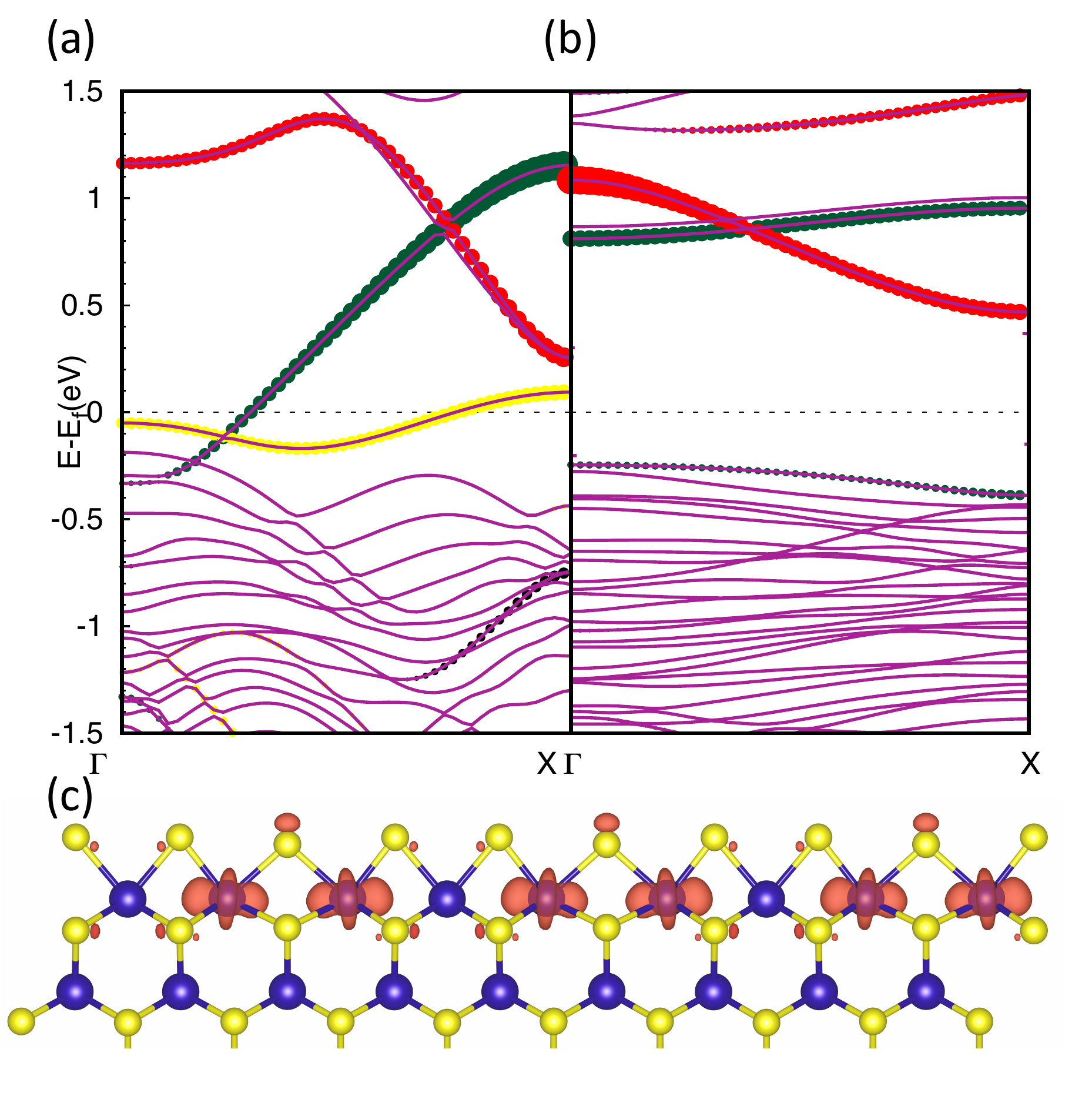}
\caption{ (a) Band structure of nanoribbon with 50\% S-dressed Mo edge in $1\times$ cell, using the same color coding as in Fig.~\ref{fig:pristineMoedge}. (b) Band structure after Peierls distortion in $3\times$ cell. (c) Charge density difference $\Delta \rho = \rho_{3\times}-\rho_{1\times}$.\cite{vesta} }
\label{fig:50dressededge}
\end{figure}

The 50\% S-dressed Mo edge has a single S attached per Mo edge atom, see Fig.\ref{fig:structures}(c). This structure emerges under growth conditions with a moderately low sulfur concentration (a moderately low sulfur chemical potential)\cite{Flemming,Lauritsen,PhysRevB.67.085410,doi:10.1021/acs.chemmater.5b00398,PhysRevB.96.165436}. The corresponding band structure of the $1\times$ cell is shown in Fig. \ref{fig:50dressededge}(a). It shows one metallic edge band, which originates from the Mo atoms at the edge. This band has the same character as the Mo $d$ band with even $\sigma_h$ symmetry identified in the pristine case, Fig. \ref{fig:pristineMoedge}(a), but is shifted downward in energy.

There is also a Mo $d$ band with odd $\sigma_h$ symmetry, with energies in the gap, which we characterized as a dangling bond state in Fig. \ref{fig:pristineMoedge}(a). In this case, however, it is completely unoccupied. Therefore, the electrons required for compensating the polarisation charge reside in the Mo $d$ band with even $\sigma_h$ symmetry. As before, topology dictates the occupation of this band to be $1/3$.
 
Reoptimizing this structure in a cell with $3\times$ periodicity, results a a sizeable Peierls distortion, where one out of the three S atoms dressing the edge moves inwards, and pushes the two Mo atoms bonded to it somewhat to the side, see Fig.\ref{fig:50dressededge}(c). In this $3\times$ pattern, two of the distances between adjacent Mo atoms along the edge then become shortened to 2.97 \AA, and the third one increases to 3.61\AA\ .  

This Peierls reconstruction opens up a band gap of $\sim 0.7$ eV, resulting in an insulating structure also for this particular Mo edge structure. The total energy decreases by a sizeable 0.33 eV per $3\times$ cell. Unlike the 100\% S-dressed Mo edge, no SDW emerges at the 50\% S-dressed Mo edge. The metal-insulator transition is Peierls driven and does not produce magnetic moments on the edge Mo atoms.   

This reconstruction agrees with results found in previous studies \cite{doi:10.1021/acs.chemmater.5b00398}. There it was also claimed that the resulting electronic structure shows a multi-valency of the Mo edge atoms, with one out of the three Mo atoms showing a higher valency than the two others. However, in our case a Bader analysis does not provide any evidence for multi-valency. 
 
\subsection{S edge} 
 

\begin {figure}[htpb]
\includegraphics[width=8.5cm]{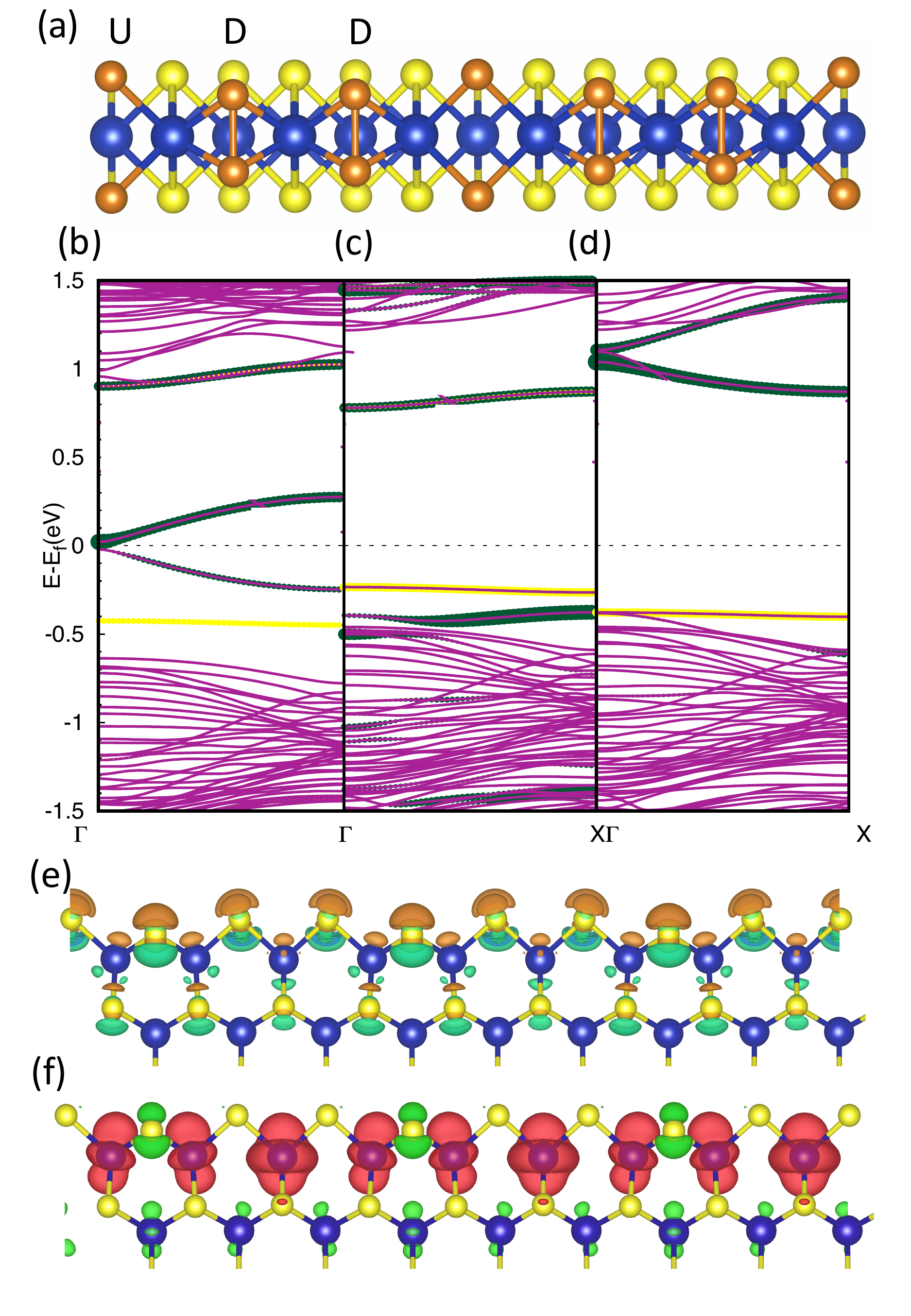}
\caption{(a) The side view of the UDD structure at the S edge, the edge S atoms are highlighted in brown (b) Band structure of nanoribbon calculated with the PBE functional, focussing on the S edge in the UDD structure; the green color measures a projection of the wave function on the Mo orbitals of even $\sigma_h$ symmetry of the Mo atoms at the S edge, the bands highlighted in yellow are the fully occupied S $p$ orbitals. (c,d) Spin-up, respectively spin-down bands in the $3\times$ periodic cell, after electronic and structural relaxation, calculated with the PBE+U functional,  $U-J=3$ eV. (e) CDW at the S edge in this cell; the brown/green colors indicate the change in charge density $\Delta \rho = \rho_{3\times}-\rho_{1\times}$ with respect to that of the unperturbed $1\times$ structure. (f) Corresponding SDW at the S edge; the red/green colors indicate the density of spin up/down wave functions\cite{vesta}. }
\label{fig:Sedge}
\end{figure}
 
The pristine S edge is terminated by S atoms, shown in Fig. \ref{fig:structures}, and has a metallic band structure dominated by an edge state that shows little dispersion, Fig. \ref{fig:pristineMoedge}(a). This state mainly consists of dangling bonds on the edge  S atoms, and is therefore susceptible to structural and chemical changes. The simplest structural change is the formation of a bond between the S edge atoms, called the dimerized (D) structure, whereas the the pristine structure is called undimerized (U). The D structure is lower in energy than the U structure by 130 meV per $1\times$ unit cell. The S dangling bond state disappears in the D structure, as expected, but the band structure is still metallic. The Fermi level is crossed by an edge state that has Mo $d$ character, dominated by the Mo atoms closest to the edge. This band is approximately $2/3$ occupied, hence it is prone to a 1D metal-insulator instability in a cell with $3\times$ periodicity. 

Turning to such a $3\times$ cell, one complicating factor is that one can have a combination D and U sulfur dimers, see Fig. \ref{fig:Sedge}(a). Using the standard PBE functional, the UUU and DDD structure are higher in energy compared to the partially dimerized and undimerized structures, UUD and UDD. The UUD structure is the lowest in energy; upon optimization it develops a CDW, which leads to opening a band gap of 0.4 eV at the edge. This is in agreement with what has been found previously for this structure \cite{doi:10.1021/acs.chemmater.5b00398}. 

However, the UDD structure shown in Fig \ref{fig:Sedge}(a) is only about 35 meV per $3\times$ unit cell higher in energy. It is still metallic with a $1/3$ occupancy of the edge state with dominant Mo $d$ character, discussed above, see Fig. \ref{fig:Sedge}(b). When on-site correlation for the Mo $d$ orbitals is turned on, with $U-J=3$ eV, the UDD structure becomes the lowest in energy compared to the UUD by almost 500 meV per $3\times$ unit cell. A strong ferromagnetic SDW develops, with magnetic moments of 1.1$ \mu_{B}$, 0.6$ \mu_{B}$, and 0.6$\mu_{B}$ on the Mo atoms at the S edge. This leads to a large band gap of $\sim 1$ eV between the spin-up bands, see Fig. \ref{fig:Sedge}(c), and an even larger band gap between the spin-down bands, Fig. \ref{fig:Sedge}(d). The SDW, shown in Fig. \ref{fig:Sedge}(f), is accompanied by a CDW, shown in Fig. \ref{fig:Sedge}(e), which also leads to a $3\times$ structural modulation. While the UDD configuration for the S atoms is retained, the distances between adjacent Mo atoms at the  edge display the $3\times$ periodicity, with values of 3.29 \AA\ and 2.98 \AA. 
 
 \section{Summary and conclusions}\label{sec:summary}
 
The physics of {\MS} edges is governed by its 2D hexagonal lattice with $D_{3h}$ symmetry, which allows for a nonzero 2D electric polarization. The latter is a topological $\mathbb{Z}_3$ invariant. It results in a polarization charge $\lambda = \pm 2e/3a$ at {\MS} zigzag edges (with $a$ the lattice constant along the edge), which needs to be compensated by an electronic charge $-\lambda$ at the edge. These electrons reside in electronic states that are confined to the edge, and have energies within the band gap of {\MS}. As these states have a total occupation of $1/3$ or $2/3$, structures with $1\times$ periodicity along the edge are necessarily metallic.
 
However, as their metallicity is one-dimensional, they are prone to periodic electronic and structural instabilities that promote a metal-insulator transition. Because of the reasoning of the previous paragraph, the periodicity of these perturbations has to be (a multiple of) three. Indeed, we find that in structures of zigzag edges discussed in this paper, instabilities of $3\times$ periodicity occur that lower the total energy, and result in a metal-insulator transition.
 
Which particular instability occurs, depends on the structure of the edge. For a Mo zigzag edge that is 50\% decorated with S atoms, a Peierls distortion occurs, which generates an insulating edge. Peierls distortions do not occur at a Mo zigzag edge that is 100\% decorated with S atoms, nor at a S zigzag edge. Instead, at these edges we find a spin-density wave (SDW), accompanied with a charge-density wave (CDW), which creates a band gap at the edges. Such SDW/CDWs have also been found at mirror-twin boundaries of {\MS} \cite{krishnamurthi2020}, indicating that they may be a common phenomenon for this material. 

In fact, because all phenomena discussed in this paper are driven by the $\mathbb{Z}_3$ topology of the 2D $D_{3h}$ symmetry, other compounds that belong to the same class as {\MS} are expected to show a similar behavior. This holds for all {\MX}, M = W, Mo; X = S, Se, Te, for instance. 

The emergence of SDW/CDWs opens up possibilities for potential experiments. The spin character might be probed by spin-polarized scanning tunneling, for instance. The $3\times$ periodicity of these SDW/CDWs also allows for interesting soliton excitations, that have fractional charges ±1/3$e$ or ±2/3$e$. Such solitons will occur naturally on zigzag edges with an overall length that is not a multiple of 3$a$. At edges with lengths that are a multiple of 3$a$, solitons do not exist in the ground state, but may be injected. In a Coulomb blockade experiment with STM, it should be possible to observe their fractional charges, for instance.
 
 \bibliography{/Users/sridevi/Desktop/Edges-Paper/2020/Jan/Version3/Edges_v3.bib} 
 
\end{document}